\documentclass[aps,prl,reprint,groupedaddress,showkeys]{revtex4-1}

\usepackage{amsmath}
\usepackage{amsfonts}
\usepackage{graphicx}
\newcommand{\poiss}[2]{\{#1,#2\}}
\begin{document}

\title{The two-dimensional three-body problem in a strong magnetic field is integrable}


\author{A. Botero}
\email[]{botero@uniandes.edu.co}
\affiliation{Departamento de F\'\i{sica, Universidad de los Andes, Bogot\'a, Colombia}}
\author{F. Leyvraz}
\email[]{leyvraz@fis.unam.mx}
\altaffiliation{Centro Internacional de Ciencias, Cuernavaca, M\'exico}
\affiliation{Instituto de Ciencias F\'\i{sicas, Universidad Nacional Aut\'onoma de M\'exico, Cuernavaca, M\'exico}}

\date{\today}

\begin{abstract}

The problem of $N$ particles interacting through pairwise  central forces
is notoriously intractable for $N\geq3$. Some quite remarkable specific
cases have been solved in one dimension, whereas higher-dimensional 
exactly solved systems involve velocity-dependent or many-body forces.
Here we show that the guiding center approximation---valid for charges moving in
two dimensions in a strong constant magnetic field---simplifies the three-body 
problem  for an arbitrary interparticle interaction invariant under rotations and
translations and makes it solvable
by quadratures. 
This includes a broad variety of special cases, such as
that of three particles interacting through arbitrary pairwise central potentials.
A spinorial representation for the system is introduced, which allows a visualization of 
its phase space as the corresponding Bloch sphere as well as the identification of a 
Berry-Hannay rotational anholonomy. Finally, a brief discussion of the quantization 
of the problem is presented.

\end{abstract}

\pacs{ab.cd}

\maketitle
It is only in a few select cases that the $N$-body problem, with $N\geq3$,
is known to be integrable.  In arbitrary dimensions, the 
best known example is that of $N$ particles interacting through linear forces, 
first solved by Newton \cite{newton}. 
In one dimension, there are several cases,
such as that of $N$ particles interacting through an $r^{-2}$ potential.
This was solved by Calogero \cite{cal69}  and Marchioro \cite{mar} and for $N=3$ 
(but see also \cite{jacobi}
for earlier related results) and by Calogero \cite{cal71} and Sutherland \cite{sut71}
for the case of the quantum system with arbitrary $N$ 
and all interaction strengths equal; the corresponding classical problem was solved by Moser \cite{mos75}.
While integrable $N$-body problems can also be found
in two and three dimensions,  these remarkable results generally involve
somewhat peculiar features, such as velocity-dependent forces, many-body interactions,
or Hamiltonians that are not of the usual form of the sum of kinetic and potential energy.
The reader will find an extensive treatment
and many references in \cite{cal01} and more recent results in \cite{cal08}. 

The aim of this letter is to present a general class of integrable systems of a rather different nature. 
On the one hand, they admit a broad 
class of interactions between the three particles: any force defined by a rotationally and translationally invariant potential
is allowed. This includes in particular the case  in which the particles interact via
arbitrary pairwise central potentials. 
On the other hand, they are explicitly limited to the case of three particles
moving in two dimensions. The feature that makes the problem solvable is that the particles are
charged with the same charge $e$ in the presence of a strong constant magnetic field $B$. The latter induces 
a rapid circular motion of particle $i$ of radius $r_i=m_iv_i/|e B|$, where $m_i$  and $v_i$,  are  the mass and 
speed of the particle respectively (we use $c=1$ throughout). If the field is sufficiently strong, the $r_i$  
become negligible relative to any other length-scales of the problem. With this fundamental assumption,  
the effective Hamiltonian system describing the secular motion becomes integrable by virtue of the 
symmetries of the interaction potential.

Let us turn to a detailed description of the system. 
Let $\vec q_i$  and $\vec p_i$ be the position and canonical momentum vectors of particle $i=1,2,3$,  with components 
$q_{i, \alpha}$ and  $p_{i, \alpha}$ respectively ($\alpha = 1,2$), and suppose the exact Hamiltonian of the system is
\begin{widetext}
\begin{equation}
H({\bf p},{\bf q})=\sum_{i=1}^3\frac1{2m_i}\left[
\left(
p_{i, 1}-e B\ q_{i,2}
\right)^2+p_{i, 2}^2
\right]+V\left(
\vec{q}_1, \vec{q}_2, \vec{q}_3
\right)+\frac{\omega_{c}}{2}\sum_{i=1}^3|\vec q_i|^2 ,
\label{eq:1}
\end{equation}
\end{widetext}
where the interaction has the symmetry
\begin{equation}
V\left(
\vec{q}_1, \vec{q}_2, \vec{q}_3
\right)=V\left(
\mathcal{R}\vec{q}_1+\vec a, \mathcal{R}\vec{q}_2+\vec a, \mathcal{R}\vec{q}_3+\vec a
\right)\, 
\label{eq:2}
\end{equation}
for arbitrary translations $\vec a$ and rotations $\mathcal{R}$ in the plane. A well-known  
transformation leads to new sets of  canonical variables: the kinematical momenta 
$\vec{\pi}_i = m \vec v_i$, and the so-called guiding centers  
$\vec R_{i} = \vec{q}_i - \hat{z} \times \vec{\pi}_i/(e B)$, which 
have the following  Poisson brackets 
\begin{subequations}
\begin{eqnarray}
\poiss{\pi_{i,\alpha}}{\pi_{i,\beta} }&=& \epsilon_{\alpha \beta}\delta_{i j} e B \label{eq:2.1a}\\
\poiss{ R_{i,\alpha}}{ R_{j,\beta} }&=& -\epsilon_{\alpha \beta}\delta_{i j} 
(e B)^{-1} \label{eq:2.1b}\\
\{ \pi_{i,\alpha}, R_{j,\beta} \}&=& 0\label{eq:2.1c}
\end{eqnarray}
\label{eq:2.1}
\end{subequations}
where $\epsilon_{\alpha \beta}$ is the antisymmetric 
tensor in two dimensions with $\epsilon_{12}=1$. As $|B|$ becomes large, the cyclotron radii  $r_i = |\vec{\pi}_i|/|e B|$ become far smaller than the scale at which the 
potential varies, and the $\vec{\pi}_i$ and $\vec{R}_i$ decouple. The guiding center motion is then 
well described by the Hamiltonian
\begin{equation}
H_{\rm gc}(\underline x, \underline y)=V\left[
(x_1,y_1), (x_2, y_2), (x_3, y_3)
\right]+\frac{\omega_{c}(|\underline x|^2+|\underline y|^2)}{2},
\label{eq:3}
\end{equation}
where the vectors $\underline x=(x_1, x_2, x_3)$ and $\underline y=(y_1, y_2, y_3)$  are the
$x$ and $y$ components of the guiding centers in units chosen so as to render them canonically conjugate: \begin{equation}
\poiss{x_i}{y_j}=\delta_{i,j},\qquad \poiss{x_i}{x_j}=\poiss{y_i}{y_j}=0.
\label{eq:4}
\end{equation}
From the Poisson brackets, it is readily seen that 
\begin{equation}
T_y=\sum_{i=1}^3x_i,\quad T_x=\sum_{i=1}^3y_i,\quad J=\frac12\sum_{i=1}^3\left(
x_i^2+y_i^2
\right),
\label{eq:5}
\end{equation}
generate translations in  $y$, translations in $x$, and rotations about the origin, all of which are symmetries of the 
interaction potential. Moreover, the harmonic external potential is proportional to $J$, which 
has vanishing Poisson bracket with the scalar 
$T_x^2+T_y^2$. We thus find two independent integrals of the motion in involution, which for later convenience, can 
be traded for functions representing  the orbital and spin angular momenta
\begin{equation}
L = \frac{1}{6}\left(T_x^2+T_y^2\right),\qquad S = J -L \, .
\label{eq:6}
\end{equation}
 Since  $L$, $S$ and the Hamiltonian  (\ref{eq:3}) are three integrals in involution, we conclude that the system is integrable.
  
 To better understand the motion in the six-dimensional phase space, we  represent the configuration of the system 
 by the triangle defined by the three particles. Due to the  harmonic potential in the Hamiltonian \ref{eq:3}, 
 the centroid coordinates $(T_y/3, T_x/3)$ of the triangle  execute uniform circular 
 motion about the origin with angular frequency $\omega_{c}$ (we take the positive sense of 
 rotation as clockwise). This motion decouples from that of the relative coordinates, which 
 describe the shape and orientation of the triangle. These can be conveniently represented in terms of the 
 {\em spinor}
\begin{equation}
\Psi=\frac{1}{2 \sqrt{3}}\left(\begin{array}{c}
{\sqrt{3}}(z_2- z_1)\\
z_2+z_1 - 2 z_3
\end{array}
\right),
\label{eq:7}
\end{equation}
where $z_i = x_i + i y_i$. As is easily verified, the spinor components satisfy the Poisson bracket relations
\begin{equation}
\poiss{\Psi_\alpha}{\Psi_\beta}=\poiss{\Psi^*_\alpha}{\Psi^*_\beta}=0,  \ \ \ \poiss{\Psi^*_\alpha}
{\Psi_\beta}=i\delta_{\alpha,\beta},\label{eq:8}
\end{equation} and have vanishing Poisson brackets with $T_x$ and $T_y$ (and hence $L$). 
The normalization of the spinor is $\Psi^\dag\Psi=S$, the conserved spin angular momentum, which 
is proportional to the square of the radius of gyration of the triangle.  A phase 
change  $\Psi \rightarrow e^{-i\chi} \Psi $ is equivalent to $(z_i - z_j) \rightarrow (z_i - z_j)e^{-i\chi}$ 
and hence to  rotating the triangle by $\chi$. Hence, the equivalence class $[ \Psi ] = \{ z \Psi | z \in \mathbb{C}\backslash\{0\} 
\}$  corresponds to
the triangle's {\em shape}, and the Bloch sphere for the unit spinors $\Psi/\sqrt S$ is the space of possible shapes.  
 \begin{figure}
\includegraphics[scale=0.5]{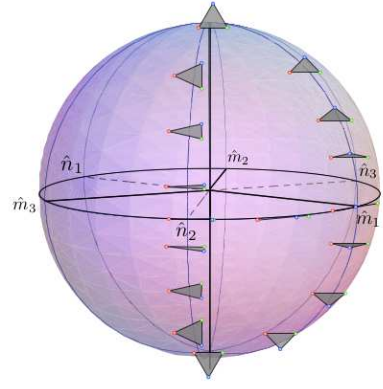}
\caption{\label{fig1}Bloch sphere representation of the triangle's shape, 
with the corresponding shapes for two meridians passing through $\hat{m}_1$ and $\hat{n}_2$. }
\end{figure}
To map shapes onto the sphere, define the unit vector 
\begin{equation}
\vec{\xi} = \frac{1}{S}\Psi^\dag \cdot\vec{\sigma} \Psi ,
\label{eq:9}
\end{equation}
where
 $\vec\sigma = (\sigma_1,\sigma_2,\sigma_3)$ is the vector of Pauli matrices \cite{text}. Now let $\rho_k$
be the squared length of the side of the triangle opposite to particle $k$ and  $A$ be the triangle's 
{\em signed area} $A=[(y_3-y_1)(x_2-x_1)-(y_2-y_1)(x_3-x_1)]/2$.  
Then these quantities are given by: 
\begin{subequations}
\label{eq:10}
\begin{eqnarray}
\rho_k&=&2S\left(1+\vec m_k\cdot\vec\xi\right)\qquad(1\leq k\leq3)\\
\label{eq:10a}
\vec m_k&=&\left(
\sin\frac{2\pi k}{3},0,\cos\frac{2\pi k}{3}
\right)
\label{eq:10b}
\\
A&=&\frac{\sqrt3}{2} S \xi_2
\label{eq:10c}
\end{eqnarray}
\end{subequations}
We can now identify special classes of triangles on the Bloch sphere (see Fig.~\ref{fig1}). Calling the intersection 
of the $y$-axis with the sphere the
North pole, and the great circle perpendicular to it the equator, we see that the poles are  
triangles of maximal area, and hence equilateral, while points on the equator, being perpendicular to $\sigma_2$,
have zero area. An isosceles triangle, say with $\rho_1=\rho_2$, is perpendicular to 
$\vec{m}_1-\vec{m}_2$, and hence lies on the great circle connecting $\vec{m}_3$ and the pole.
Therefore, the three 
great circles  through the poles and the $\vec{m}_k$ describe isosceles triangles. The 
$\vec{m}_k$ and their antipodes--- denoted by $\vec{n}_k$---are thus simultaneously 
isosceles and {\em collinear\/} triangles, the $\vec m_k$ with two identical  sides adding up to the longest side, 
the  $\vec{n}_k$  
with two coincident vertices. 

In these new coordinates, the Hamiltonian is given by
\begin{equation}
H_{\text rel}=V\left(\Psi^\dag\Psi, \Psi^\dag\cdot\vec\sigma\Psi\right)+\omega_{c}  \Psi^\dag\Psi
=V\big(S,\vec\xi\big)+\omega_c\xi,
\label{eq:11}
\end{equation}
where $S$ and $\vec \xi$ are as defined earlier. From the Poisson 
brackets (\ref{eq:8}) and Hamilton's equations $\dot{\Psi}_\alpha =\{ \Psi_\alpha, H \}$, we find 
that the spinor satisfies the equation of motion
\begin{equation}
i \dot{\Psi} = \left[\omega_c +  \left(\frac{\partial V}{\partial S}\right)_{\vec \xi} + 
\frac{1}{S}
\sum_{k=1}^3\left(
\frac{\partial V}{\partial \xi_k}\right)_S\left(
\sigma_k-\xi_k\,\openone
\right)
\right] \Psi \, ,
\label{eq:12}
\end{equation}
where the subscripts in the derivatives indicate fixed variables. The equation of motion in shape space is then 
\begin{equation}
\dot{\vec \xi} =\frac{2}{S}  \left(\nabla_{\vec \xi}V \right)_S \times \vec \xi \, ,
\label{eq:13}
\end{equation}
tracing orbits along the level curves of $V$ on the Bloch sphere. Note that for homogeneous 
$V$ of degree $\lambda$, $V( a S,\vec \xi )= a^\lambda V(S,\vec \xi )$, the shape dynamics 
becomes independent of $S$ when expressed in scaled time $\tilde{t} = S^{\lambda-1} t$. 

So far we have looked at the dynamics of the rotation invariant 
characteristics of the triangle, namely $S$ and $\vec\xi$. Analysis of the rotational motion is more subtle. 
While triangle rotations correspond to phase changes $\Psi\rightarrow e^{-i\chi} \Psi $, the
notion of overall phase for $\Psi$ is ill-defined. Still, a
relative phase $d\chi = i  \Psi^{\dag} d \Psi/(\Psi^{\dag} \Psi)$ 
can be defined between infinitesimally separated spinors $\Psi$ and $\Psi + d \Psi$. Hence, a 
{\em dynamical} angular velocity can be defined to account for infinitesimal changes in orientation:
\begin{equation}
\omega_r^{(dyn)}= \frac{1}{S}\Psi^\dag \dot{\Psi} = \omega_c + 
\left(\frac{\partial V}{\partial S}\right)_{\vec \xi} 
\, .
\label{eq:14}
\end{equation}
Remarkably, when $V$ is homogeneous of degree $\lambda$,  
$\omega_r^{(dyn)}=\omega_{c}+\lambda V_0/S $, 
where $V_0$ is the conserved potential energy.  For finite times, the rotation 
angle is only well defined if the initial and final shapes are equal, that is, over a period $T_s$ 
of the shape motion. One would then expect that 
$\Delta \chi = \int_{0}^{T_s}\!  \omega_r^{(dyn)} dt $ is the net rotation of the triangle, but this is incorrect. 
The reason is that  an additional  {\em geometric} phase, or Berry--Hannay phase 
\cite{be84,han85,wil89}, is acquired by parallel transport.  Explicitly, if $\Psi$ is parameterized  as 
\begin{equation}
 \Psi= \sqrt{S}\,e^{-i \gamma }\left(\begin{array}{c}
\cos \frac{\theta}{2}\\
\sin \frac{\theta}{2}e^{- i \phi }
\end{array}
\right), 
\label{eq:15}
\end{equation}
we find that
$
 \omega_r^{(dyn)} = \dot{\gamma} + \sin^2 \frac{\theta}{2} \dot{\phi} \, .
$
So the spinor  acquires a phase $\Delta \gamma$, translating to the angular velocity $\omega_r = \frac{\Delta \gamma}{T_s}$:
\begin{equation}
\omega_r = \langle \omega_r^{(dyn)} \rangle + \omega_r^{(geo)}, \qquad \omega_r^{(geo)}\equiv - \frac{\Omega(E,S)
}{2 T_s}\,
\label{eq:16}
\end{equation}
where $\langle \omega_r^{(dyn)} \rangle$ is the time average of $\omega_r^{(dyn)}$ 
in the period $T_s$ and  
$\Omega(E,S) = 2\oint d\phi \sin^2 \frac{\theta}{2}$ is the oriented solid angle on the Bloch 
sphere enclosed by the  level curve of $E$, as follows from Stoke's theorem.  As expected, 
$\omega_r =  \omega_r^{(dyn)} $ at a fixed point of the shape motion. 

Further analysis of the shape and rotational dynamics is possible in terms of action-angle variables, the 
details of which lie beyond the scope of this paper. Still, some salient features are worth mentioning. 
For the rotational motion, $S$ is the natural action variable. The  action variable $I_s$ for the shape 
motion is obtained from the integral   $\frac{1}{2 \pi} \oint \underline{x} \cdot d \underline{y}$ \cite{arn97}, 
around the closed circuit in phase space given by a level curve of the energy on the Bloch sphere with 
fixed centroid. The final result has a nice geometric interpretation, namely
\begin{equation}
I_s = S \frac{\Omega(E,S)}{4 \pi} \, .
\label{eq:17}
\end{equation}
This relation can be used to show that $T_s$, the characteristic period
for the shape motion, is given by 
$T_s= \frac{S}{2} \frac{\partial \Omega(E,S)}{\partial E}$, and that $\omega_r$ in 
(\ref{eq:16})
is indeed the characteristic frequency for the rotational motion. 
 \begin{figure}
\includegraphics[scale=0.45]{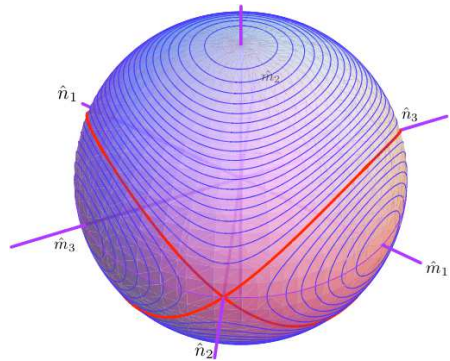}
\caption{\label{fig2}Bloch sphere phase portrait for an interparticle central potential $u(r) \propto r^6$.}
\end{figure}

To fix ideas further, it may be useful to describe a special case in greater detail. Thus, if all three particles
are identical and interact pairwise through a central potential of the form 
$u(r) \propto r^{2 \lambda}=\rho^\lambda$, then one finds generically a phase portrait on the Bloch sphere
with 8 critical points (Fig. \ref{fig2}), two of which are the north and south poles, which are always stable equilibria. The remaining 
six points are the $\vec{m}_k$ and $\vec{n}_k$ on the equator. For $\lambda>2$, the $\vec{m}_k$ are stable (elliptical) fixed points 
while the $\vec{n}_k$ are unstable (hyperbolic) fixed points; 
for $\lambda<2$ with $\lambda \neq 1$, on the other hand, the $\vec{m}_k$ are unstable (hyperbolic) 
fixed points, whereas  the $\vec{n}_k$
are either stable fixed points ($0 < \lambda < 2$) or points at which the energy diverges ($\lambda <0$); in the latter case nearby 
orbits encircle these points as if they were elliptical fixed points, and describe systems in which two particles  
revolve rapidly around each other in a tightly bound orbit, with a distant third particle slightly
perturbing the motion, in a pattern somewhat analogous to that of the Sun--Earth--Moon system.  The two 
exceptions to this pattern are worthy of mention: First is the case $\lambda=1$ ($u(r) \propto r^2 $) 
 for which $V \propto S$, so the triangle shape is frozen and $\omega_r $ is constant. The other 
 exception is $\lambda=2$ ($u(r) \propto r^4 $), in which case $V \propto 6 S^2 (3 - \cos^2 \theta)$ 
 where $\theta$ is the polar angle. In this case both the area and 
the radius of gyration of the triangle are constants of the motion, and for $\omega_c=0$ both $\omega_r $ and $\omega_s$ scale 
linearly with this area, with $\omega_s/\omega_r$ depends on  $\theta$ and is independent of the size
 $S$. 

We conclude with a brief discussion of the quantum mechanics of the problem. Re-interpreting the guiding center 
variables $\{x_i \}$ and $\{y_i\}$ as canonical operators with the Poisson brackets in Eq.~(\ref{eq:4}) replaced  
by commutators (times $i$), we define the annihilation operators $a_i = \frac{1}{\sqrt{2}}(x_i + i y_i)$, so that 
$[a_i,a_j^\dag] = \delta_{ij}$. To separate centroid and relative motion,  define the  operators
\begin{equation}
\left(\begin{array}{c} b \\ \Psi_1 \\ \Psi_2 \end{array}\right) = 
\left(\begin{array}{c c c} 
\frac{1}{\sqrt{3}} & \frac{1}{\sqrt{3} }& \frac{1}{\sqrt{3}}\\ 
-\frac{1}{\sqrt{2}} & \frac{1}{\sqrt{2}} & 0 \\
\frac{1}{\sqrt{6}} & \frac{1}{\sqrt{6} }& -\frac{2}{\sqrt{6}}\end{array}\right)\left(\begin{array}{c} a_1 \\ a_2 \\ a_3 \end{array}\right)
\label{eq:18}
\end{equation}
and their adjoints, with $[b,b^\dag] = [\Psi_\alpha,\Psi_{\alpha}^\dag] = 1$ and all other 
commutators vanishing;  in analogy with the
classical case, the $\Psi_\alpha$ are interpreted as components of an operator-valued 
spinor. In terms of these, the orbital and spin angular momentum operators are  
\begin{equation}
L = b^\dag b+ \frac{1}{2}, \qquad S = \Psi^\dag \Psi + 1,
\label{eq:19}
\end{equation}
with eigenvalues that are positive half-integers or integers respectively. The $\Psi_\alpha$  implement 
a Schwinger oscillator construction \cite{schwinger} of an $SU(2)$ algebra for the shape description.  To see this, define
\begin{equation}
F_i = \frac{1}{2} \Psi^\dag\cdot\sigma_ i \Psi,
\label{eq:20}
\end{equation}
in terms of which the squared inter-particle distances are $\rho_k = 2 S + 4 \vec F \cdot \vec{m}_k$ and the triangle 
area is $A = \sqrt{3} F_2$. The $F_i$ satisfy the commutation relations $[F_j,F_k] = i \epsilon_{jkl}F_l$, and thus 
commute with $F^{2} =\vec F \cdot \vec{F}$, which is found to be
\begin{equation}
F^{2} = \frac{S^{2} -1}{4} \, .
\label{eq:21}
\end{equation}
Therefore each eigenspace of $S$ with eigenvalue $s$ defines an $s$-dimensional spin-$(s-1)/2$ irreducible 
representation of $SU(2)$ for the  $F_i$. With the previous symmetries of the interaction potential, the guiding center 
Hamiltonian can be written as $H_{\text{gc}} = V(S, \vec F) + \omega_c (L + S)$. The eigenvalues can then be labelled 
by the three quantum numbers of the problem: $ l , s$ and $n$, with $E_{l,s,n} = \tilde{E}^{(s)}_{n} + \omega_c(l + s)$, 
where the $\tilde{E}^{(s)}_{n}$ are the eigenvalues of the shape Hamiltonian for each sector $s$, namely, the reduced 
$s\times s$ matrix obtained from the potential with the $F_i$ replaced by their respective 
representation matrices ${F_i}^{(s)}$:
\begin{equation}
H^{(s)} = V(s, \vec{F}^{(s)}).
\label{eq:22}
\end{equation}
Finally, to account for particle statistics, we note that $H^{(s)}$ can further be block-diagonalized into sectors transforming  irreducibly under the action of the permutation group $S_3$, from which the relevant symmetric or antisymmetric subspaces can be identified.

\begin{acknowledgments}
FL gratefully acknowledges funding provided by 
UNAM DGAPA--PAPIIT grant number IN114014 as well as CONACyT grant 
154586. AB gratefully acknowledges funding by Uniandes, proyecto No. 114-2013. 
\end{acknowledgments}


\end{document}